\documentclass[aps,12pt,showpacs]{revtex4}
\usepackage{amssymb} %
\usepackage{graphicx,setspace}
\usepackage{graphicx} %
\usepackage{amsmath,euscript}
\usepackage{color}
\newcommand{\be}{\begin{equation}}
\newcommand{\ee}{\end{equation}}
\newcommand{\prt}{\partial}
\newcommand{\sech}{\mathop{\rm
sech}\nolimits} %

\begin{document} %

\title{Influence of shear-flow vorticity on wave-current
interaction.\protect \\ Part 1: Surface gravity waves without
surface tension effect.} %
\author{Philippe Ma{\"{\i}}ssa$^{1)}$, Germain Rousseaux$^{2*)}$,
Yury Stepanyants$^{3)}$} %
\affiliation{$^{1)}$ Universit\'{e} de Nice-Sophia Antipolis,
Laboratoire J.-A. Dieudonn\'{e}, UMR CNRS-UNS 7351, Parc Valrose,
06108 Nice Cedex 02, France, European Union; \\%
$^{2)}$ Institut Pprime,  UPR 3346, CNRS  -- Universit\'{e} de
Poitiers -- ISAE  ENSMA, 11 Bd Marie et Pierre Curie, BP 30179,
86962 Futuroscope, France, European Union; \\%
$^{3)}$ Faculty of Health, Engineering and Sciences, University of
Southern Queensland, Toowoomba, QLD, 4350, Australia.}

\date{\today}

\begin{abstract}%
\vspace*{2cm} Propagation of surface waves on a
background shear flow with constant vorticity is studied and
compared against the case when the background flow is uniform in
depth. For a shear flow with the linear vertical profile, the
dispersion relation of surface gravity waves is minutely analyzed
for a fluid of finite depth. Under the assumption that the
background flow gradually varies in the horizontal direction, the
primary attention is paid to the wave blocking phenomenon; the
effect of vorticity on this phenomenon is studied in details. The
conditions of wave blocking are obtained and categorized for
different values of the governing dimensionless parameters.
\end{abstract}


\maketitle

\vfill

\noindent
-----------------------------------\\
{\small *) Corresponding author; e-mail:
germain.rousseaux@univ-poitiers.fr

\clearpage

\section{Introduction}
\label{Intro}

In many cases water waves propagate in a moving water which leads
to the complex phenomena of wave-current interaction. Usually,
currents are nonuniform both in the vertical and horizontal
directions. Many examples of wave propagation on the background of
spatially nonuniform shear flows may be listed in the conjunction
with this phenomenon. Among them there are tidal currents,
currents caused by the wind stress, oceanic currents of different
scales, currents induced by long surface or internal waves, river
flows, etc. It is well-known that the interaction of water waves
with nonuniform currents may result in either dramatic effects
such as generation of freak waves \cite{Igor1, KharPelSlun-09} or
favorable effects such as the blocking of waves by pneumatic
wave-breakers \cite{Taylor}.

Due to the effect of the bottom friction at the sea bed and/or of
wind stress at the free surface, the currents often vary with the
depth. The result of such vertical variation is the appearance of
shear-flow vorticity which plays an important role in the wave
variability and stability. Moreover, as has been shown in
\cite{Shrira-89}, the vorticity can even lead to the existence of
new specific type of waves -- the vorticity waves. Thus, the
shear-flow vorticity in combination with the intensity of a
current introduces an important parameter which should be taken
into account in the attempts to understand the physics of the
wave-current interaction. Leaving aside numerous works on the
influence of shear flows on the stability of internal waves in
density stratified fluid, we only refer here to some papers where
the interaction of surface waves with currents has been studied
(there is actually a plethora of publications on this problem; it
is impossible to refer to all of them, therefore we restrict
ourselves by references to the papers which are the most relevant
to our work): \cite{Thompson, Biesel, Burns, Tsao, Fenton, Brevik,
Skop, Kirby, Makarova, Nepf, Kantardgi, Helena, Choi}. Here, we
revisit this problem following the recently developed dynamical
system approach \cite{PRL09, NJP10}. We focus on the simplest case
of a shear flow with the linear vertical profile having constant
vorticity; more general case will be treated elsewhere. We show
that the problem can be analyzed in terms of dimensionless
parameters and introduce the `effective' Froude number defined
such that the flow vorticity is taken into account along with the
speed of a current.

In this paper we do not consider the effect of surface tension.
This issue is very interesting and important, but is much more
complicated. It will be analyzed separately in the Part II, which
will be published later.

\section{The blocking phenomenon of surface gravity waves
on the current slowly varying in horizontal direction}%
\label{Sect1}%

Let us consider one-dimensional wave propagation on a surface of
moving water of a finite depth $h$. The velocity profile of the
water flow $U(z)$ may be a rather complicated function of the
depth, but in this paper we focus on the consideration and
intercomparison of two cases, i) when the flow is uniform, $U(z) =
U_0$ and ii) when it is a linear function of depth $U(z) = U_0 +
\alpha z$. Here $U_0$ is the water speed at the free surface, and
$\alpha$ characterizes the vorticity of the background flow. It is
assumed that the axis $z$ is directed upward with zero at the
unperturbed water surface. For certainty we suppose that $U_0
> 0$, i.e. we assume that the background flow is co-directed
with the axis $x$. The sketch of the considering flow is shown in
Fig.~\ref{f01}. %

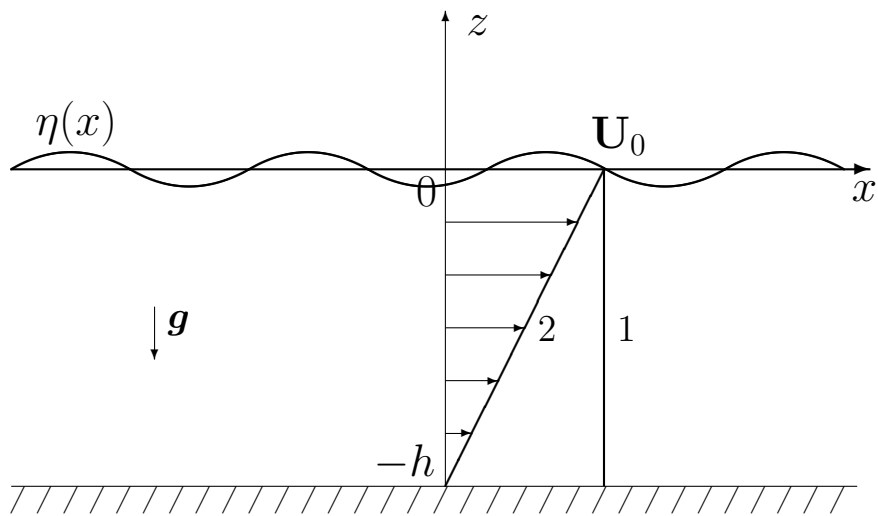
\begin{figure}[h]
\vspace{7cm}%
\begin{picture}(300,6)%
\put(170,2){\vector(0,1){180}}%
{\thicklines %
\put(170,2){\line(1,2){60}}%
\put(230,2){\line(0,1){120}}%
}%
\put(170,102){\vector(1,0){50}}%
\put(170,82){\vector(1,0){40}}%
\put(170,62){\vector(1,0){30}}%
\put(170,42){\vector(1,0){20}}%
\put(170,22){\vector(1,0){10}}%
{\thicklines %
\hspace*{-5mm}
\put(20,122){\vector(1,0){325}}%
\put(338,110){{\Large $x$}}%
\qbezier(20,122)(42,135)(65,122)%
\qbezier(65,122)(87,109)(110,122)%
\qbezier(110,122)(132,135)(155,122)%
\qbezier(155,122)(177,109)(200,122)%
\qbezier(200,122)(222,135)(245,122)%
\qbezier(245,122)(267,109)(290,122)%
\qbezier(290,122)(312,135)(335,122)%
}%
\put(20,2){\line(1,0){320}}%
\multiput(25,2)(10,0){32}{\line(-1,-2){5}}%
\hspace*{5mm}
\put(178,173){{\Large $z$}}%
\put(159,108){\Large $0$}%
\put(143,6){{\Large $-h$}}%
\put(65,62){{\large $\boldsymbol{g}$}}%
\put(60,70){\vector(0,-1){20}}%
\put(225,130){{\Large \bf{U}$_0$}}%
\put(15,135){{\Large $\eta(x)$}}%
\put(235,57){{\large 1}}%
\put(205,57){{\large 2}}%
\end{picture}
\vspace{5mm} %
\caption{Sketch of the fluid flow in the reference coordinate
frame associated with the immovable bottom. Line 1 depicts the
velocity with the uniform profile and line 2 -- the velocity with
the linear profile.}%
\label{f01}%
\end{figure}

In the general case, when the shear velocity is uniform in the
horizontal direction, but is an arbitrary function of $z$, the
dispersion relation between the frequency $\omega$ and the wave
number $k$ for waves of infinitesimal amplitude can be derived in
the approximate form based on the Taylor series representation
(see, e.g., \cite{Skop,Kirby,Swan-James-2001}). However, recently
it was found \cite{Karageorgis2012} that in many particular cases
the dispersion relation can be presented in the explicit
analytical form. Two simple cases of analytical representation of
the dispersion relation in the closed form are very well known
(see below), they are those which have been mentioned above and
shown in Fig.~\ref{f01}, i.e. when the velocity profile is either
depth independent or varies linearly with the depth (in the latter
case the flow vorticity is constant). These two profiles were
chosen for the analysis of the vorticity influence on the
wave-current interaction and comparison of results obtained. Note
that when the fluid velocity vanishes at the bottom (see
Fig.~\ref{f01}), then $\alpha = U_0/h$, but in general $\alpha$
may be considered as the independent parameter characterizing the
flow vorticity.

The dispersion relation for surface gravity waves in a water with
the linearly-varying current has been derived by Thompson and
Biesel \cite{Thompson, Biesel}:
\begin{equation}%
\label{DispRel}%
\omega^{\pm}(k) = U_0 k - \frac{\alpha}{2}\tanh{kh} \pm
\sqrt{\left(\frac{\alpha}{2}\tanh{kh}\right)^2 + gk\tanh{kh}} \, ,%
\end{equation} %
where $g$ is the acceleration due to gravity.

The dispersion relation (\ref{DispRel}) formally consists of four
branches corresponding to different choices of signs of $\omega$
and $k$ and describing co- and counter-propagating waves with
respect to the background flow. Clearly, there is a certain
symmetry around the origin of the $(k, \omega)$ coordinate frame,
so that for a given set of parameters $U_0$, $\alpha$ and $h$, we
have $\omega^+(k)= - \omega^-(-k)$, and the phase velocity $c
\equiv \omega^+(k)/k = \omega^-(-k)/(-k)$. As the physical
frequency is the non-negative quantity ($\omega = 2\pi/T$, where
$T$ is the wave period), therefore we will consider further only
non-negative values of the dispersion relation $\omega^+(k)$ for
all possible values of $k$ varying from minus to plus infinity. At
such conventional agreement, $k > 0$ pertains to co-propagating
surface waves with $c \ge 0$, and $k < 0$ pertains to
counter-propagating surface waves with $c \le 0$.

Note that under such conventional agreement the wave frequency in
the frame moving with the certain speed $c_0 = U_0[1 -
\tanh{kh}/(2kh)]$ is non-negative for all possible wave numbers
$k$: $\omega - U_0k + \alpha\tanh{kh}/2 \ge 0$. Meanwhile, in the
immovable coordinate system, the frequency formally may become
negative for certain negative values of $k$. In such case we will
consider, in accordance with the aforementioned symmetry, another
branch of the dispersion relation with $\omega > 0$ and $k > 0$;
this will be clarified and illustrated below in Fig.~\ref{f02}.

By introducing the normalized variables, $\kappa = kh$ and
$\tilde\omega = \omega h/U_0$, Eq.~(\ref{DispRel}) can be
presented in the dimensionless form
\begin{equation}%
\label{DLDisp} %
\tilde\omega(\kappa) = \kappa - \frac{\Omega}{2}\tanh{\kappa} +
\sqrt{\left(\frac{\Omega}{2}\tanh{\kappa}\right)^2 + \left[\left(1
- \frac{\Omega}{2}\right)^2 - \left(\frac{\Omega
\mbox{F}_\Omega}{2}\right)^2\right]\frac{\kappa\tanh{\kappa}}{\mbox{F}_\Omega^2}}
\, ,
\end{equation}%
where $\Omega = \alpha h/U_0$ is the dimensionless parameter
characterizing the vorticity of the basic flow (the actual
vorticity is $-\Omega$), and $\mbox{F}_\Omega = (1 -
\Omega/2)(\Omega^2/4 + gh/U_0^2)^{-1/2}$ is the `effective' Froude
number defined such that the flow vorticity is taken into account.
If $\Omega = 0$, the background flow is uniform in depth with no
vorticity, and the effective Froude number reduces to the
conventional Froude number $\mbox{F}_\Omega = \mbox{Fr} \equiv
U_0/\sqrt{gh}$. If $\Omega = 1$ the background flow has a linear
profile vanishing at the bottom as shown in Fig.~\ref{f01} (the
constant vorticity flow), then $\mbox{F}_\Omega =
\mbox{Fr}/\sqrt{4 + \mbox{Fr}^2}$. The effectiveness of
introduction of the new form of the Froude number is especially
clear when we consider the long-wave limit of the dispersion
relation (\ref{DispRel}):
\begin{equation}%
\label{LWDispRel}%
\omega^{\pm}(k)|_{k \to 0} = \left(U_0 - \frac{\alpha
h}{2}\right)k \pm
k\sqrt{\left(\frac{\alpha h}{2}\right)^2 + gh} \, .%
\end{equation} %
In this case one can introduce the `effective' uniform velocity
$\bar U = U_0 - \alpha h/2$ and speed of long waves $\bar c =
\sqrt{\left(\alpha h/2\right)^2 + gh}$. Then the Froude number can
be defined in the traditional manner:
\begin{equation}%
\label{NewFroude}%
\mbox{F}_\Omega = \frac{\bar U}{\bar c} = \frac{U_0 - \alpha
h/2}{\sqrt{\left(\alpha h/2\right)^2 + gh}} = \frac{1 -
\Omega/2}{\sqrt{\Omega^2/4 + gh/U_0^2}}. %
\end{equation}

Consider now the case when the current is uniform in depth
($\Omega = 0$) and present the dispersion relation (\ref{DLDisp})
graphically (see Fig.~\ref{f02}). The co-propagating waves in
accordance with our assumption that $U_0
>0$ are those waves whose phase velocities $\tilde c \equiv
\tilde\omega/\kappa$ are positive, i.e. the corresponding
dispersion curves are located to the right from the vertical axis
in Fig.~\ref{f02}. The situation is reversed for the
counter-propagating waves; the corresponding dispersion curves are
located to the left from the vertical axis. (Note that the group
velocity, $\tilde c_g \equiv d\tilde\omega/d\kappa$, may be
oppositely directed with respect to the phase velocity; we will
address to this issue hereafter.)

\begin{figure}[!ht] %
\begin{center} %
\includegraphics[width=16cm]{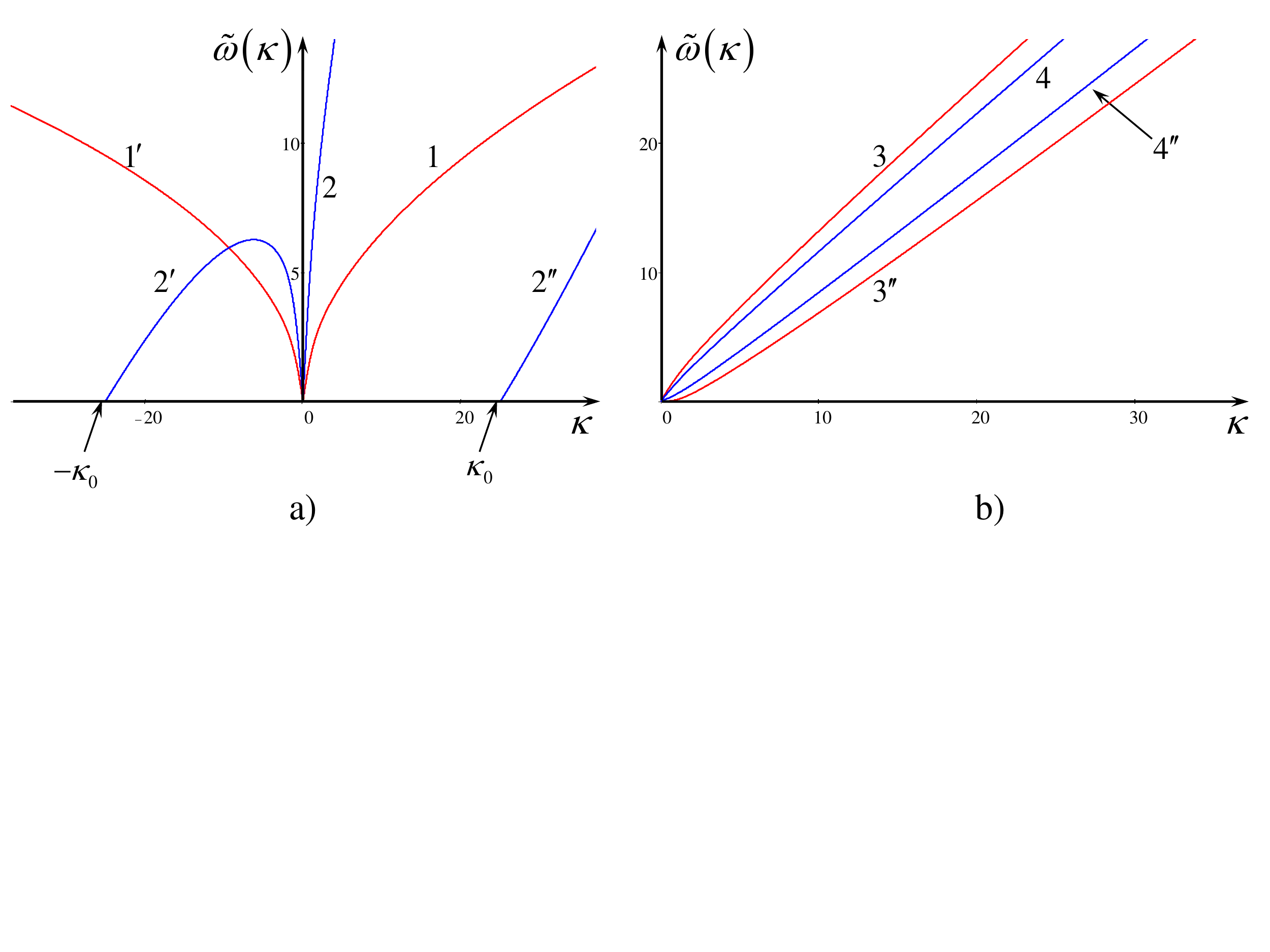} %
\vspace*{-5cm} %
\caption{(color online) Dimensionless frequency of small amplitude
water waves against the dimensionless wave number as per
Eq.~(\ref{DLDisp}) with $\Omega = 0$. Lines 1 and $1'$ pertain to
Fr = 0.01, lines 2, $2'$ and $2''$ pertain to Fr = 0.2, lines 3
and $3''$ pertain to Fr = 1, and lines 4 and $4''$ pertain to Fr =
2. Lines 1 and $1'$ are 50 times compressed in the
vertical direction. Scales in the frames a) and b) are different.}%
\label{f02}%
\end{center}%
\end{figure}%

When the Froude number goes to zero (i.e. when the current
velocity vanishes), then two branches of the dispersion curve look
almost symmetrical (cf. lines $1$ and $1'$ in Fig.~\ref{f02}a).
Asymptotically when $\kappa \to 0$, the dispersion relation
(\ref{DLDisp}) reduces to $\tilde\omega =
|\kappa|\left(1/\mbox{Fr} + \kappa/|\kappa|\right)$. The
dispersion curves become more and more asymmetrical as the Froude
number increases.

At a certain value of $\kappa = -\kappa_0$, the left branch of the
dispersion curve turns to zero (see line $2'$ in Fig.~\ref{f02})
and becomes formally negative for $\kappa < -\kappa_0$. In
accordance with the aforementioned symmetry,  this line reappears
as the line $2''$ in the right half-plane for $\kappa \ge
\kappa_0$. So, the gravity water waves with sufficiently large
wavelengths whose wave numbers occupy the range $-\kappa_0 <
\kappa < 0$ can counter-propagate against the current, whereas
shorter gravity waves cannot propagate against the flow and they
are simply pulled down by the current. The energy of such waves
which are pulled down by the background current is negative in the
immovable coordinate system \cite{Nezlin-76, OstrRybTsim-86,
StepFabr-89, FabrStep-98}. As shown in the Appendix, the wave
energy $E$ is proportional both to the frequency $\tilde\omega$
(which may be {\it formally} of either sign) and the relative
frequency $\tilde\omega - \kappa$ (which is the frequency in the
frame co-moving with the fluid and is always non-negative). The
waves with negative energy are potentially unstable and may grow
if there is a mechanism taking away their energy (their negative
energy becomes even ``more negative'' in this case, and the wave
amplitude increases in the result). There are several different
mechanisms of energy removal from the waves leading to various
types of shear flow instabilities (e.g., the dissipative
instability, radiative instability, and so on), for more details
see \cite{Nezlin-76, OstrRybTsim-86, StepFabr-89, FabrStep-98}.

The condition for the existence of negative-energy waves (NEWs)
can be found from the dispersion relation (\ref{DLDisp}) (for
details see Appendix): the NEWs are waves whose frequency
$\tilde\omega$ is {\it formally} negative, but replaced by
positive value with the corresponding exchange $-\kappa \to
\kappa$. In the dispersion plane of Fig.~\ref{f02} the NEWs are
shown by typical lines $2'', 3''$ and $4''$. Correspondingly,
waves with the positive frequency $\tilde\omega$ possess positive
energy (positive-energy waves -- PEWs). The NEWs may exist only in
a moving fluid for any value of Froude number if the surface
tension effect is neglected.

The boundary between the PEWs and NEWs can be readily found in
general for $\Omega \ne 0$ from the condition $\tilde\omega = 0$,
which gives the critical value of wave number $\kappa_0$ for given
parameters $\mbox{F}_\Omega$ and $\Omega$:
\begin{equation}%
\label{kappa0} %
\frac{\kappa_0}{\tanh{\kappa_0}} = \Omega + \frac{1}{\mbox{Fr}^2}
\equiv \Omega\left(1 - \frac{\Omega}{4}\right)\left(1 -
\frac{1}{\mbox{F}_{\Omega}^2}\right) +
\frac{1}{\mbox{F}_{\Omega}^2}.
\end{equation}%

In the limiting cases of uniform flow ($\Omega = 0$) and a shear
flow with constant vorticity like in Fig.~\ref{f01}) ($\Omega =
1$), Eq.~(\ref{kappa0}) reduces to
\begin{eqnarray}%
\frac{\kappa_0}{\tanh{\kappa_0}} &=& \frac{1}{\mbox{Fr}^2} \quad
(\mbox{recall that } \mbox{Fr} \equiv \mbox{F}_\Omega \mbox{ when
} \Omega = 0); \nonumber \\
{} && \nonumber \\
\frac{\kappa_0}{\tanh{\kappa_0}} &=& \frac{1}{4}\left(3 +
\frac{1}{\mbox{F}_{\Omega}^2}\right), \quad \Omega = 1. \nonumber
\end{eqnarray}

In Appendix we show that the same expression follows from the
direct calculation of wave energy from the linearised set of
hydrodynamic equations [cf. Eq.~(\ref{A15})].

The analysis of Eq.~(\ref{kappa0}) shows that the roots of this
equation may exist only when $\mbox{F}_\Omega < 1$ (see
Fig.~\ref{f03}). When $\mbox{F}_\Omega = 1$, the horizontal line 2
in Fig.~\ref{f03} touches line 1 at its minimum; the branch
corresponding to PEWs disappears. And, at last, when
$\mbox{F}_\Omega > 1$, Eq.~(\ref{kappa0}) does not have real
roots, and surface waves with any wave number become NEWs
including infinitely long waves with $\kappa = 0$.
\begin{figure}[!htbp]%
\begin{center}%
\includegraphics[width=12cm]{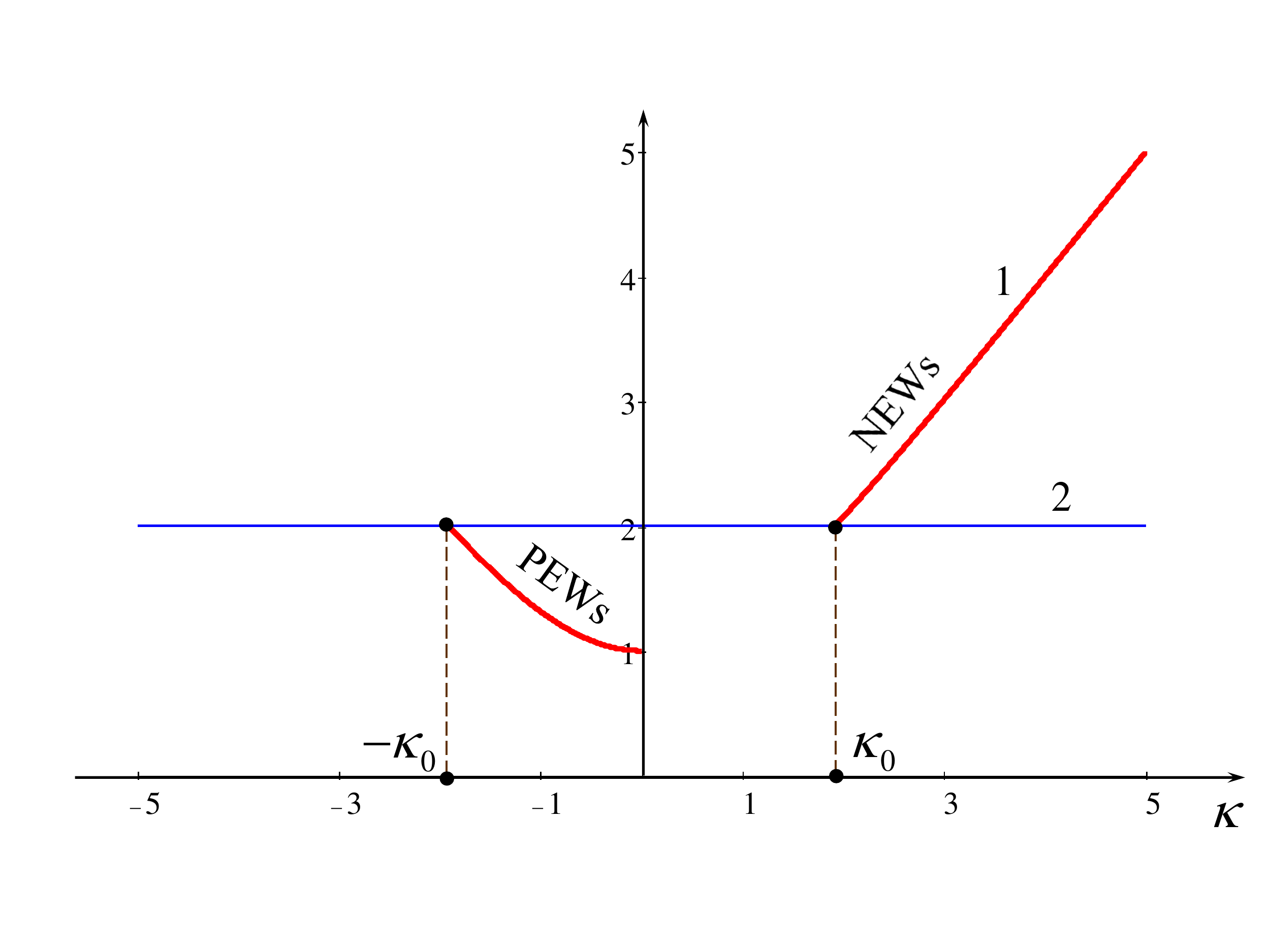}%
\vspace*{-0.5cm} %
\caption{(color online) Graphical solution of Eq.~(\ref{kappa0}):
line 1 represents the function $\kappa/\tanh{\kappa}$, whereas the
horizontal line 2 represents the right-hand side of
Eq.~(\ref{kappa0}). NEWs exist above line 2 ($\kappa > \kappa_0$),
whereas PEWs -- below that line $-\kappa_0 <
\kappa < 0$.} %
\label{f03}%
\end{center}%
\end{figure}%

When the right-hand side of Eq.~(\ref{kappa0}) increases, the
value of the parameter $\kappa_0$ also increases. However, when
$\kappa$ becomes large enough, other factors, such as the surface
tension may come into play and modify Eq.~(\ref{kappa0}); we do
not consider such effects in this paper leaving them for a
separate publication. When the right-hand side of
Eq.~(\ref{kappa0}) decreases and approaches unity, the parameter
$\kappa_0$ vanishes, which implies that gravity waves of any
wavelengths including infinitely very long waves cannot propagate
against the current any more; the current is so strong that it
pulls down even very long waves.

In what follows, we will consider the propagation of
quasi-sinusoidal wave train with a fixed frequency assuming that
the current velocity and vorticity smoothly vary in horizontal
direction so that the characteristic scale of current variation is
{\it much greater} than the characteristic wavelength of the wave
train. In such case, the frequency of the wave train remains
constant, whereas its wave number adiabatically varies in space in
accordance with the flow variation. To determine the character of
space variation of the carrier wavenumber we need to present the
dispersion equation (\ref{DLDisp}) in terms of
$\kappa[\tilde\omega, \mbox{F}_\Omega(x)]$ assuming that the wave
frequency is fixed. Unfortunately, such solution cannot be
presented in the explicit analytical form in general for the
arbitrary depth; however it can be obtained in the limiting cases
of deep and shallow water when $\vert\kappa\vert \gg 1$ or
$\vert\kappa\vert \ll 1$, correspondingly. Nevertheless, some
interesting features of a wave motion can be derived even in the
general case of a fluid of arbitrary constant depth.

\subsection{Wave blocking in the general case of arbitrary water depth}%
\label{Subsect1}%

If the wave train counter propagates with respect to the gradually
increasing main current, the group velocity of such wave train
decreases. At a certain point the wave train may stop if its group
speed $\tilde c_g \equiv d\tilde\omega/d\kappa$ turns to zero (for
the chosen fixed wave frequency, the dispersion relation has no
more real solution for $\kappa$). In this case the ``{\it wave
blocking phenomenon}'' occurs. There is a plethora of papers
devoted to this phenomenon; it is not possible to list all of them
in this paper, therefore we refer only to some of them:
\cite{Peregrine, Hedges, Jonsson, Thomas, Dingemans, Nezlin-76,
OstrRybTsim-86, StepFabr-89, FabrStep-98, Igor1, Thompson, Biesel,
Burns, Tsao, Fenton, Brevik, Skop, Kirby, Makarova, Nepf,
Kantardgi, Helena, Choi}.

Figure \ref{f04} illustrates schematically the blocking phenomenon
with subsequent generation of reflected waves at the blocking
point. In the geometry considered here, the incident wave denoted
by 2 has positive energy and negative both group and phase
velocities; it propagates to the left. All reflected waves have
positive group velocities, but their phase velocities are
different. For the first of them denoted by $2'$ the phase
velocity is negative, whereas for two other reflected waves 1 and
$2''$ the phase velocities are positive. The energies of reflected
waves are positive and negative as indicated in Fig.~\ref{f04}.
Here the schematic picture of wave transformation is presented for
pure gravity surface waves; it becomes much more complicated when
the surface tension effect is taken into consideration. We plan to
present our analysis with the surface tension effect in a separate
publication.

\begin{figure}[!htbp]%
\begin{center}%
\includegraphics[width=12cm]{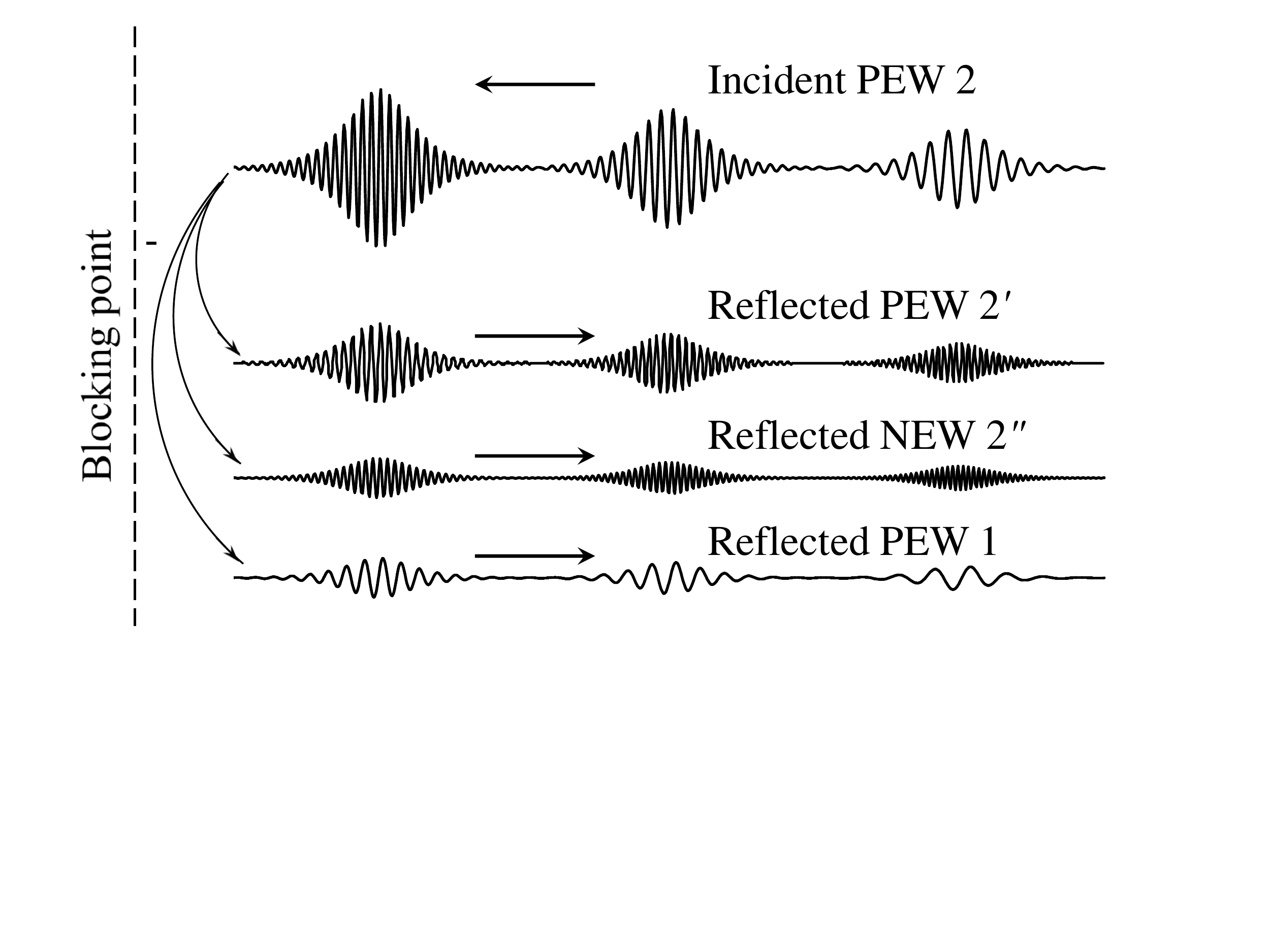} %
\vspace*{-2.5cm}%
\caption{Schematic illustration of wave blocking phenomenon for
pure gravity waves. The incident wave gives rise to three
reflected waves in general. The plots are not in scale, therefore
variations of amplitude and wave number within each wave train is
just qualitative.}
\label{f04}%
\end{center}%
\end{figure}

As has been mentioned above, the condition for the wave blocking
phenomenon is the vanishing of the group speed of a wavetrain.
Differentiating the dispersion relation (\ref{DLDisp}) for
counter-propagating waves over $\kappa$, the following equation
can be derived:
$$
\Omega^2\left\{\left(1 -
\frac{\kappa}{\tanh{\kappa}}\right)\left(1 -
\Omega\sech^2{\kappa}\right) + \left[4\sech^2{\kappa} - \left(1 +
\frac{2\kappa}{\sinh{2\kappa}}\right)^2
\right]\frac{\Omega^2}{16}\right\}\mbox{F}^4_\Omega + {}%
$$
$$
\left(2 -
\Omega\right)^2\left\{\frac{\kappa}{\tanh{\kappa}}\left(1 -
\Omega\sech^2{\kappa}\right) - \left[2\sech^2{\kappa} - \left(1 +
\frac{2\kappa}{\sinh{2\kappa}}\right)^2
\right]\frac{\Omega^2}{8}\right\}\mbox{F}^2_\Omega - {}%
$$
\begin{equation}%
\label{TranscEq}%
\frac{\left(2 - \Omega\right)^4}{16}\left(1 +
\frac{2\kappa}{\sinh{2\kappa}}\right)^2 = 0.
\end{equation}

The critical wave number corresponding to the blocking phenomenon
follows from this transcendental equation at the given values of
$\mbox{F}_\Omega$ and $\Omega$. However, the equation is algebraic
(bi-quadratic) with respect to $\mbox{F}_\Omega$ and by solving it
one can find the Froude number $(\mbox{F}_\Omega)_b$ at which all
wavetrains having carrier wavenumber greater than $\kappa_b$ are
blocked, where $\kappa_b$ is the root of Eq.~(\ref{TranscEq}). Let
us analyze Eq.~(\ref{TranscEq}) in the limiting cases of $\Omega =
0$ and $\Omega = 1$.

There is only one root of Eq.~(\ref{TranscEq}) when the flow is
uniform, i.e. when $\Omega = 0$:
\begin{equation}%
\label{OneRoot}%
\mbox{F}_\Omega\left(\kappa_b, 0\right) = \frac{2\kappa_b +
\sinh{2\kappa_b}}
{2\sinh{2\kappa_b}}\sqrt{\frac{\tanh{\kappa_b}}{\kappa_b}}.%
\end{equation}
And there are two roots for $\mbox{F}_\Omega^2$ when $\Omega = 1$,
but only one of them has the physical meaning:
\begin{equation}%
\label{FrSq2}%
\mbox{F}_\Omega^2\left(\kappa_b, 1\right) = \frac{4\kappa^2 -
8\sinh^2{\kappa}\left(1 + \cosh{2\kappa}\sqrt{1 +
4\kappa^2}\right) + \sinh^2{2\kappa} +
4\kappa\sinh{2\kappa}\left(1 + 4\sinh^2{\kappa}\right)}{4\kappa^2
+ \sinh^2{2\kappa} - 64\sinh^4{\kappa} + 4\kappa\sinh{2\kappa} +
16\sinh^2{\kappa}\left(2\kappa\sinh{2\kappa} - 1\right)}.%
\end{equation}%
Asymptotically at $\kappa_b \to -\infty$ this dependences reduce
to
$$
\mbox{F}_\Omega\left(\kappa_b, 0\right) \approx
\frac{1}{2\sqrt{-\kappa_b}}, \quad \mbox{F}_\Omega\left(\kappa_b,
1\right) \approx \frac{1}{4\sqrt{-\kappa_b}}.
$$

The relationships between $\left(\mbox{F}_\Omega\right)_{b} \equiv
\mbox{F}_\Omega(\kappa_b)$ and $\kappa_b$ as per
Eqs.~(\ref{OneRoot}) and (\ref{FrSq2}) are single-valued; they are
shown in Fig.~\ref{f05}. It follows from this figure in accordance
with Fig.~\ref{f03} that all counter-propagating waves including
infinitely long waves are blocked if
$\left(\mbox{F}_\Omega\right)_{b} \ge 1$.

Eliminating $\mbox{F}_{\Omega}$ from the dispersion relation
(\ref{DLDisp}), one finds the frequency of the wavetrain at the
blocking point $\kappa = \kappa_b$; the dispersion curve
$\tilde\omega(\kappa)$ has a local maximum at this point (see line
$2'$ in Fig.~\ref{f02}). For $\Omega = 0$ and $\Omega = 1$ one can
readily find
\begin{eqnarray}%
\tilde\omega_{b}(\kappa_b,0) &=& \kappa_b -
\frac{2\kappa_b\sinh{2\kappa_b}}
{2\kappa_b + \sinh{2\kappa_b}}; \label{FreqMax0}\\%
&& \nonumber \\
\tilde\omega_{b}(\kappa_b,1) &=& \kappa_b -
\frac{\tanh{\kappa_b}}{2}\left[1 + \sqrt{1 +
\frac{4\kappa_b\tanh{2\kappa_b}}{\sech{2\kappa_b} -
2\kappa_b\tanh{2\kappa_b}
+ \sqrt{1 + 4\kappa_b^2}}}\,\right]. \label{FreqMaxPN}%
\end{eqnarray}%

Combination of Eqs.~(\ref{OneRoot}) and (\ref{FreqMax0}) or
Eqs.~(\ref{FrSq2}) and (\ref{FreqMaxPN}) gives the parametric
representation of blocking wave frequency $\tilde\omega_{b}$ on
the critical Froude number $\left(\mbox{F}_\Omega\right)_b$. The
corresponding dependences are shown in Fig.~\ref{f06}a) for
$\Omega = 0$ and $1$.

\begin{figure}[!t]%
\begin{center}%
\includegraphics[width=12cm]{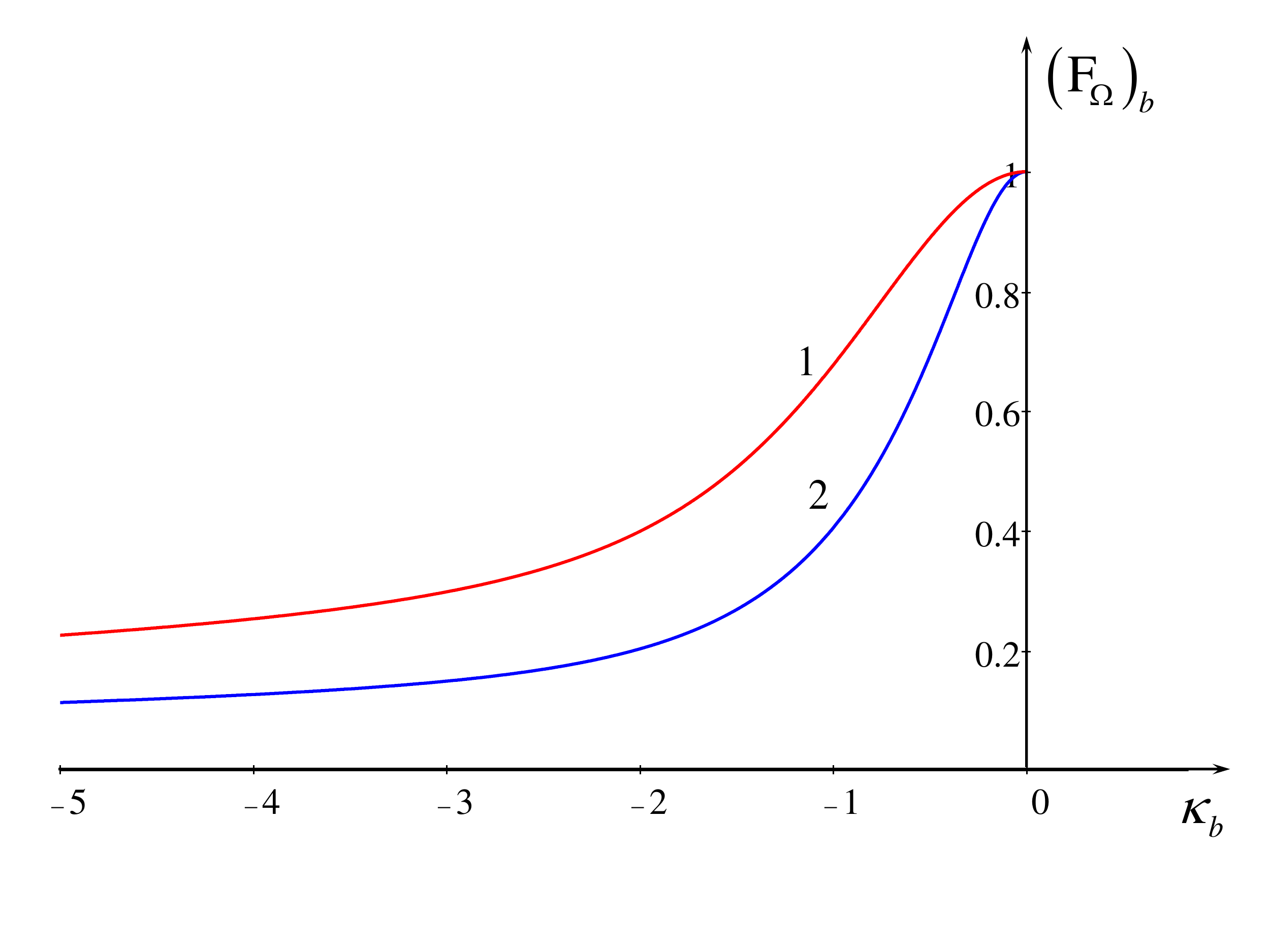}
\vspace*{-0.5cm}%
\caption{Dependences $\mbox{F}_{\Omega}(\kappa_b)$ as per
Eqs.~(\ref{OneRoot}) and (\ref{FrSq2}) for $\Omega = 0$
(line 1) and $\Omega = 1$ (line 2).}%
\label{f05}%
\end{center}%
\end{figure}

The relationship between the frequency of the blocked wave and
Froude number can be also presented in terms of the dependence of
critical Froude number on the normalized wave period $\tilde T_b
\equiv T_b\sqrt{g/h}$. It can be presented again in the parametric
form, where $\kappa$ plays the role of the parameter:
\be%
\label{UvsT-ParamPos}%
\left(\mbox{F}_{\Omega}\right)_b = \mbox{F}_{\Omega}(\kappa_b,
\Omega), \qquad \tilde T_b = \frac{\pi}{\tilde
\omega_b}\sqrt{\left[\frac{2 -
\Omega}{\left(\mbox{F}_\Omega\right)_b}\right]^2 - \Omega^2},
\ee%
where Eqs.~(\ref{OneRoot}) and (\ref{FreqMax0}) should be used for
$\Omega = 0$, and Eqs.~(\ref{FrSq2}) and (\ref{FreqMaxPN}) for
$\Omega = 1$. These dependences are shown in Fig.~\ref{f06}b).

\begin{figure}[!b]%
\includegraphics[width=13cm]{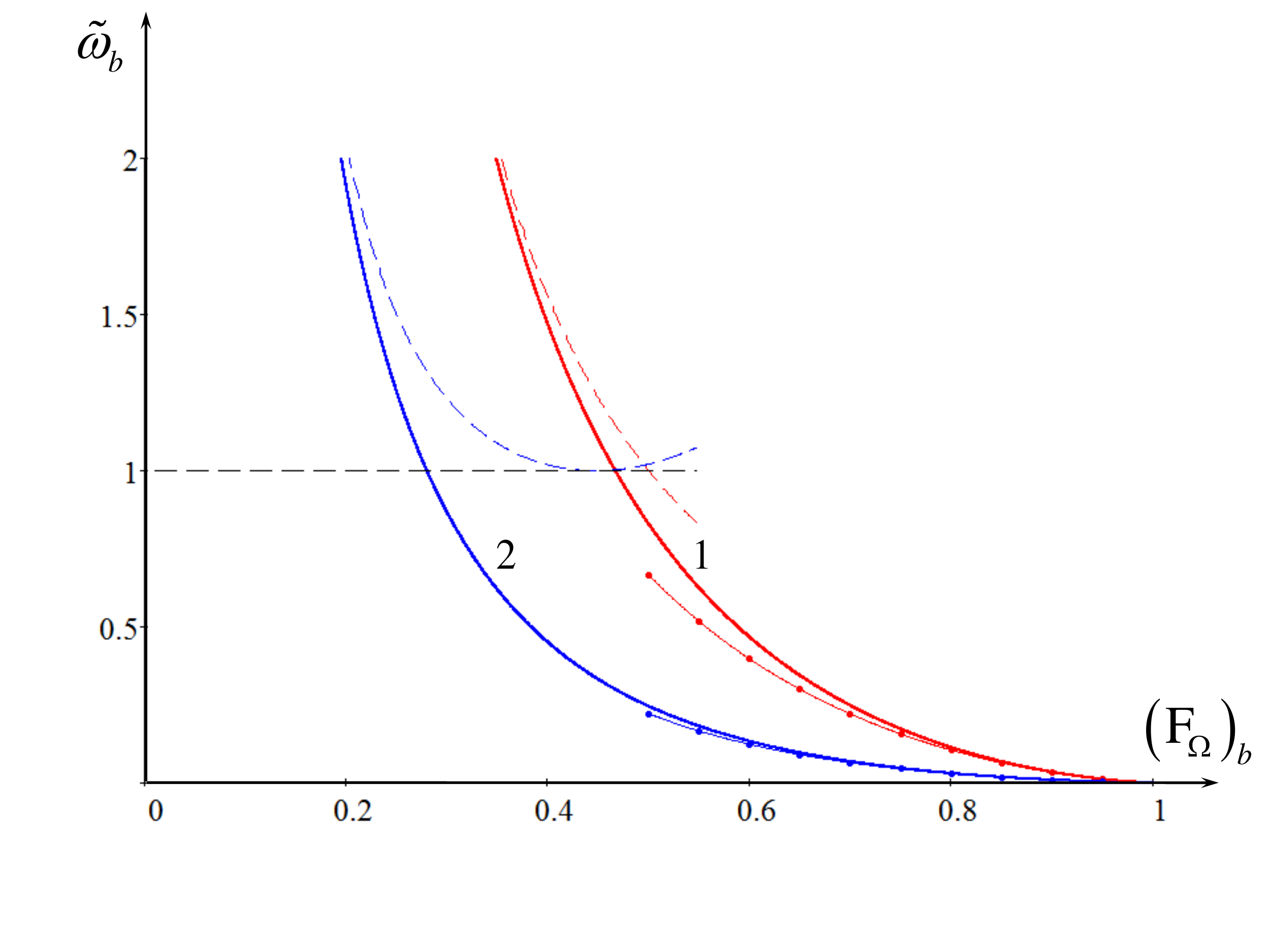} %
\vspace*{-1.0cm}%
\hspace*{1cm}%
\includegraphics[width=13cm]{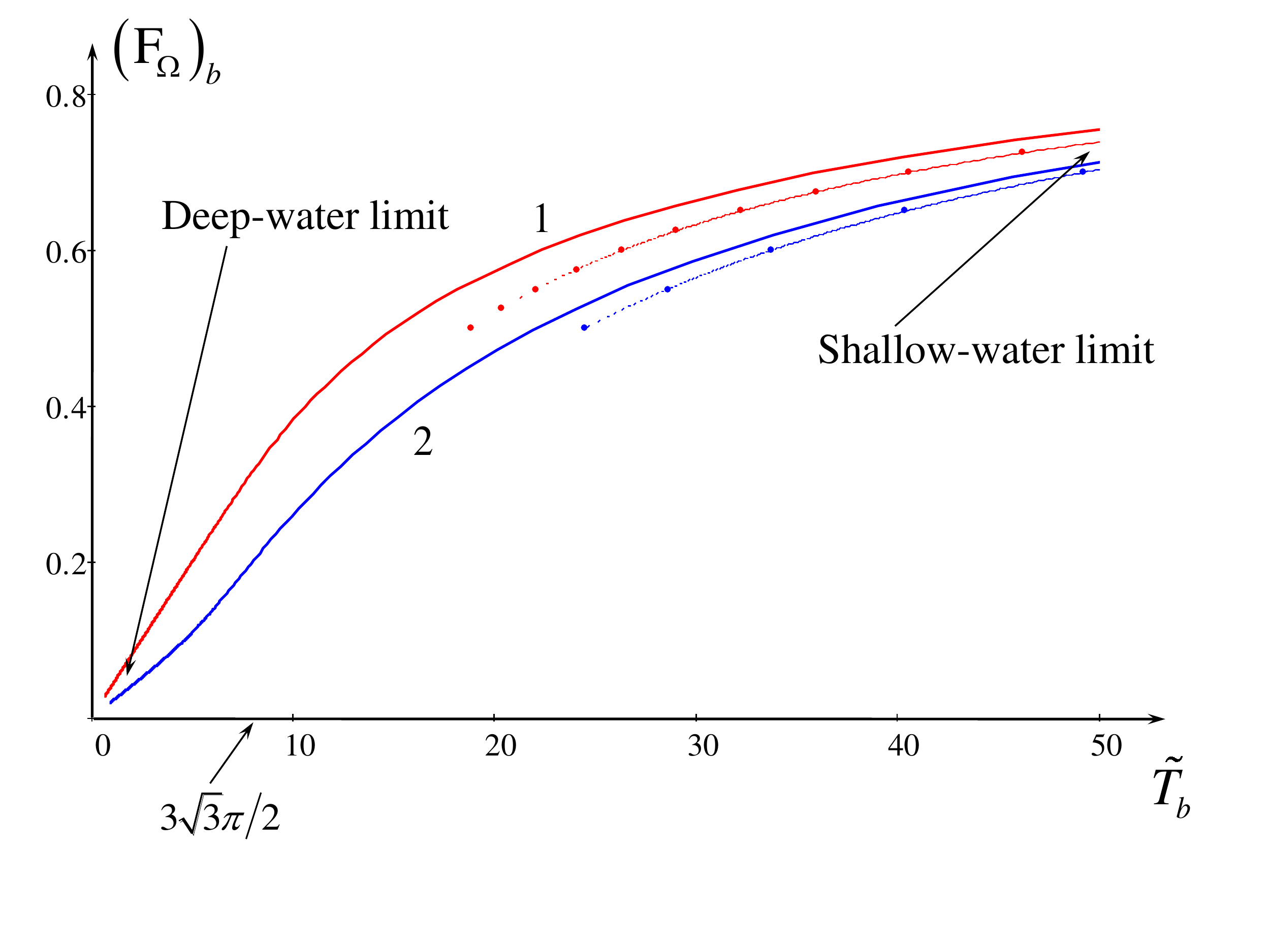} %
\begin{center}%
\vspace*{-1.0cm} %
\begin{picture}(300,6)%
\put(-50,400){\large a)}%
\put(-50,150){\large b)}%
\end{picture}
\caption{a) Dependences of dimensionless frequency
$\tilde\omega_{b}$ on the critical Froude number. \\
\protect\phantom{Figureb,}b) Critical Froude number against the
carrier wave period $\tilde
T_b$. \\
Lines 1 pertain to $\Omega = 0$, and lines 2 -- to $\Omega = 1$.
Dashed lines show the corresponding dependences for the deep-water
case, whereas dashed-dotted lines show the dependences for the
shallow-water case (see below).}%
\label{f06} %
\end{center} %
\end{figure}

In the limiting cases of deep and shallow water the corresponding
dependences can be derived in the explicit forms (see the
following subsections).

\subsection{The deep-water case}%
\label{Subsect2}%

Within the framework of the deep-water approximation ($\kappa <
0$, but $|\kappa| \gg 1$ the dispersion relation (\ref{DLDisp})
can be represented as the quadratic polynomial of $\kappa$:
\begin{equation}%
\label{DWDLDisp} %
\kappa^2 + \left[2\tilde\omega  - \Omega - \frac{\left(2 -
\Omega\right)^2}{4\mbox{F}_\Omega^2} +
\frac{\Omega^2}{4}\right]|\kappa| + \tilde\omega(\tilde\omega -
\Omega) = 0 \, .%
\end{equation}
Here for the characteristic spatial scale one can use the scale of
variation of the background flow with depth $h$ or, if the flow is
uniform in depth, the initial value of the carrier wave length
$\lambda_0$ chosen at a certain reference point.

As has been aforementioned, the condition for the wave blocking
phenomenon is the vanishing of the group speed of a wavetrain.
This corresponds to the situation when the discriminant of
Eq.~(\ref{DWDLDisp}) is zero, and two roots of this equation
becomes equal \cite{PRL09, NJP10}. Figure \ref{f08} illustrates
such situation for $\Omega = 0$ (frame a) and $\Omega = 1$ (frame
b). Line $g$ pertains to the greater root $\kappa_g(\tilde\omega)$
of Eq.~(\ref{DWDLDisp}), and line $s$ pertains to the smaller root
$\kappa_s(\tilde\omega)$ of the equation. Note that
Fig.~\ref{f08}a) is just another representation of Fig.~\ref{f02},
but in the deep-water limit; lines $g'$ and $s'$ in
Fig.~\ref{f08}a) correspond to line $2'$ in Fig.~\ref{f02},
whereas lines $s$ and $g$ correspond to lines 2 and $2''$,
respectively.

\begin{figure}[!h]%
\begin{center}%
\includegraphics[width=8cm]{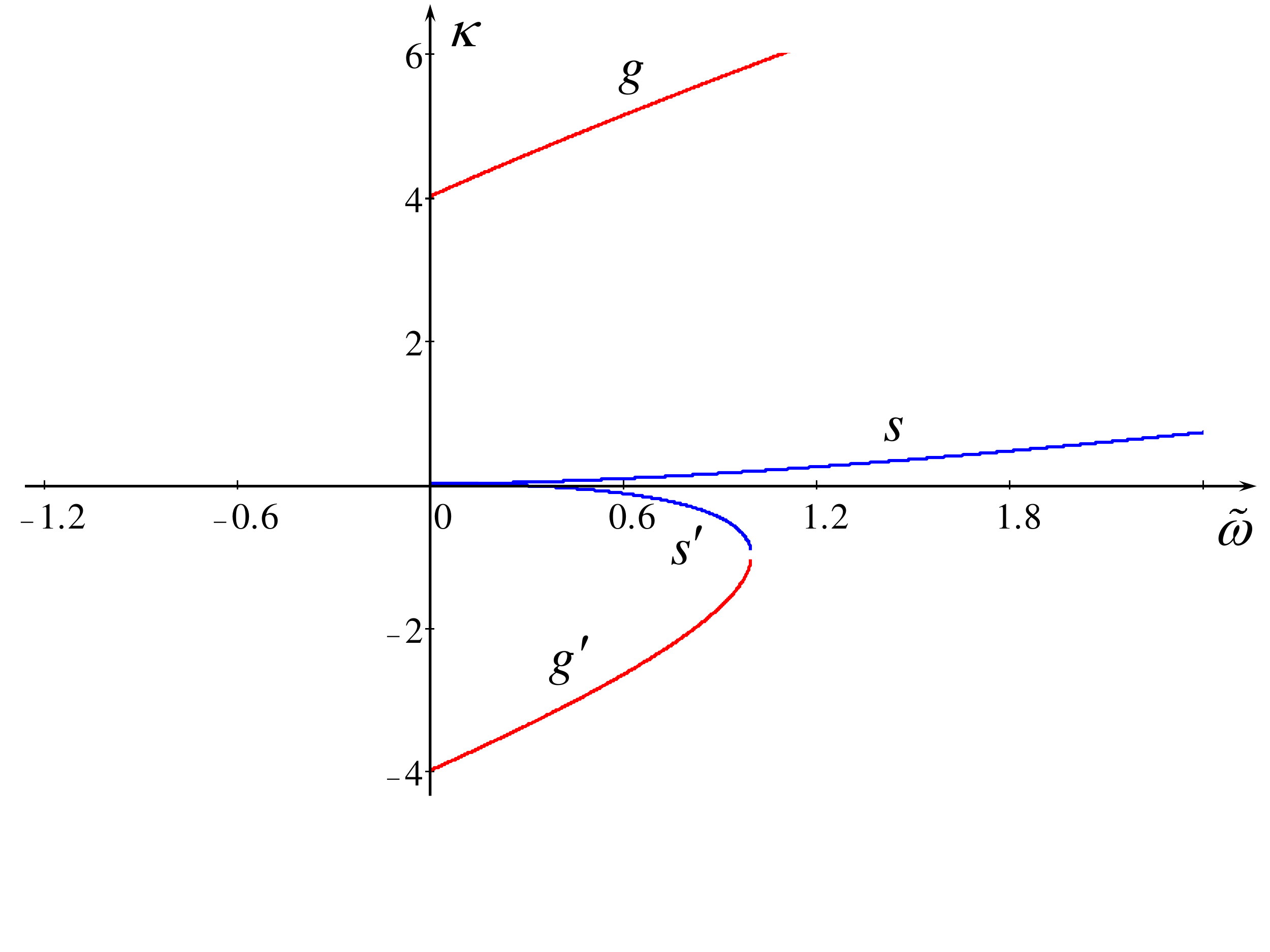}%
\includegraphics[width=8cm]{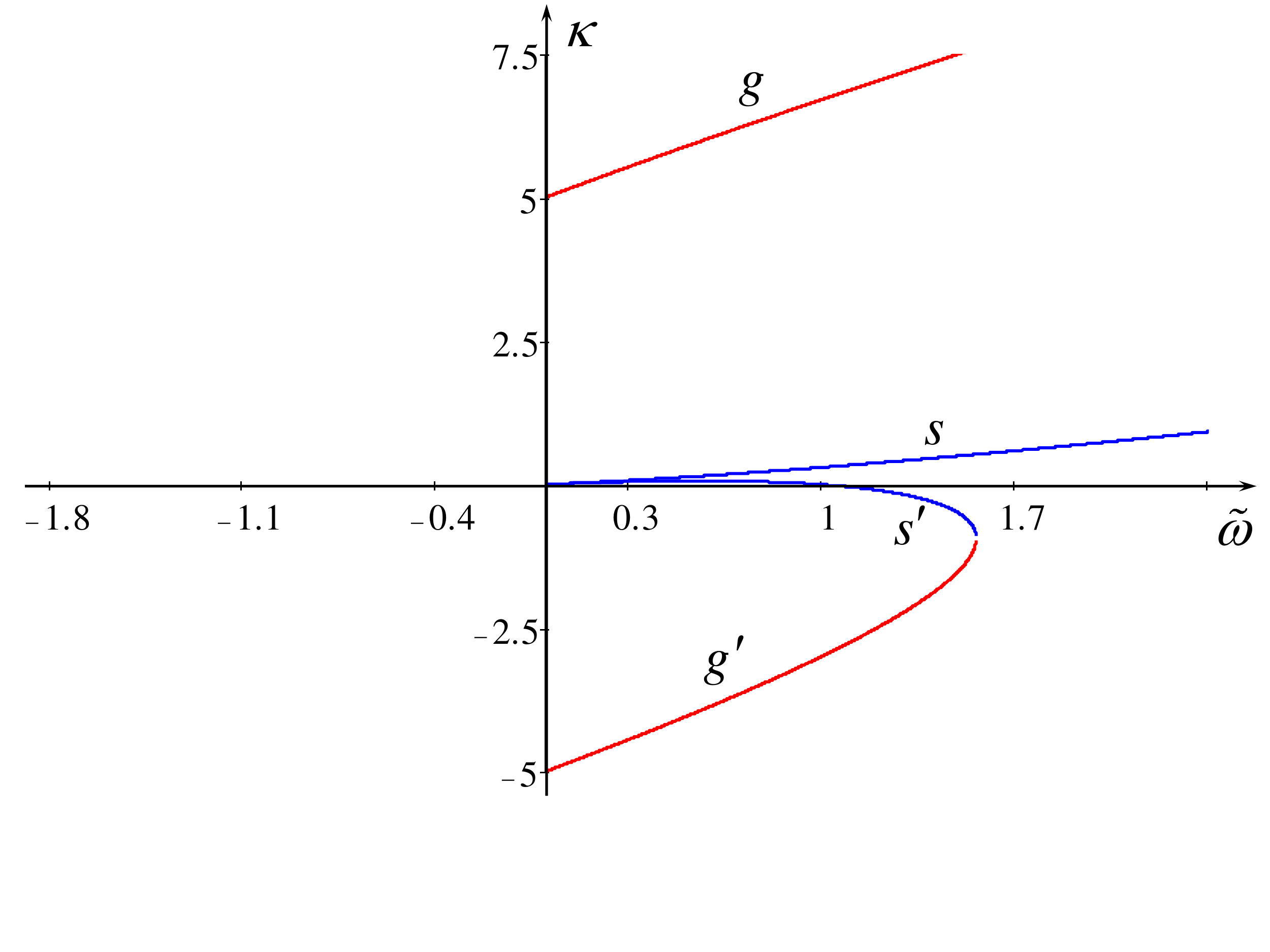}%
\vspace*{0.5cm} %
\caption{Wave number versus frequency as per Eq.~(\ref{DWDLDisp})
for the particular values of $\mbox{F}_\Omega \equiv \mbox{Fr} =
0.5$ and $\Omega = 0$ (frame a) and 1 (frame b). Portions of
curves $g$ and $s$ which correspond to $\tilde\omega < 0$ (they
are shown by dashed lines) are redrawn for $\tilde\omega > 0$
(lines $g'$ and $s'$), but for $\kappa < 0$ (for explanation see
the paragraph after Eq.~(\ref{DispRel})).}
\begin{picture}(300,6)%
\put(-5,90){\large a)}%
\put(250,90){\large b)}%
\put(59,170){\circle*{3}}%
\put(61,162){$A$}%
\put(327,172){\circle*{3}}%
\put(329,164){$A$}%
\end{picture}
\label{f08}%
\end{center}%
\end{figure}
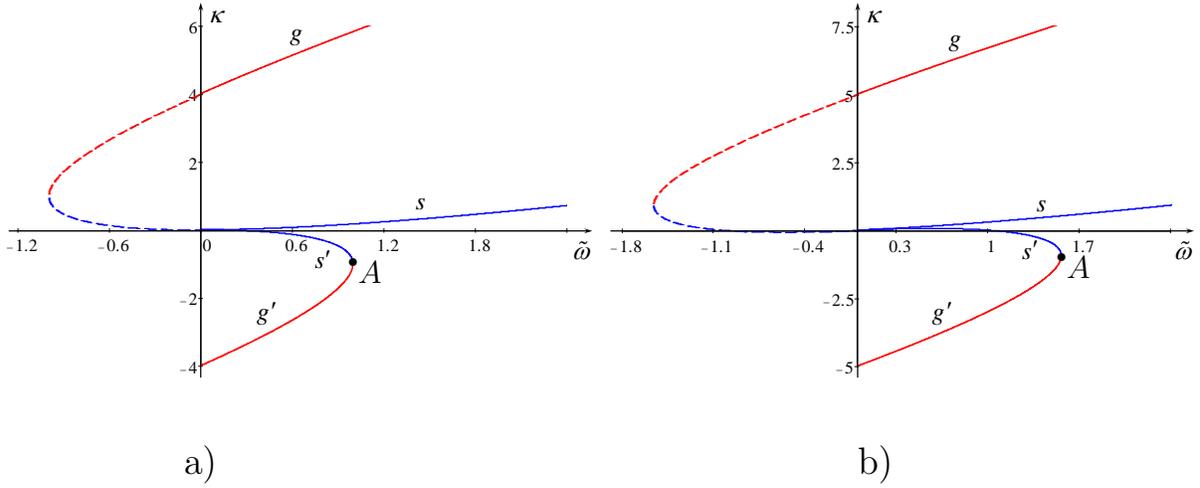


When the discriminant of Eq.~(\ref{DWDLDisp}) is zero (this
corresponds to points $A$  shown in Fig.~\ref{f08}), the
dispersion relation can be presented in the factorized form:
\begin{equation}%
\label{DoubleRoot}%
(\kappa - \kappa_b)^2 = 0, \quad \mbox{where} \quad \kappa_b =
\tilde\omega  - \frac{\Omega}{2} - \frac{\left(2 -
\Omega\right)^2}{8\mbox{F}_\Omega^2} + \frac{\Omega^2}{8}.
\end{equation}%

The corresponding value of the frequency when the discriminant
vanishes is:

\begin{equation}%
\label{CondDoubleRoot}%
\tilde\omega_{b} =
\frac{1}{16\left(\mbox{F}_\Omega\right)_b^2}\frac{\left\{4 -
\Omega(4 - \Omega)\left[1 -
\left(\mbox{F}_\Omega\right)_b^2\right]\right\}^2}{\left\{4 -
\Omega(4 - \Omega)\left[1 -
\left(\mbox{F}_\Omega\right)_b^2\right]\right\} -
4\Omega\left(\mbox{F}_\Omega\right)_b^2}.
\end{equation}%
In the particular cases, $\Omega = 0$ and $\Omega = 1$, this
equation reduces to

\begin{equation}%
\label{DoubRtPart}%
\tilde\omega_{b}|_{\Omega = 0} =
\frac{1}{4\left(\mbox{F}_\Omega\right)_b^2}, \quad
\tilde\omega_{b}|_{\Omega = 1} =
\frac{1}{16\left(\mbox{F}_\Omega\right)_b^2}\frac{\left[1 +
3\left(\mbox{F}_\Omega\right)_b^2\right]^2}{1 -
\left(\mbox{F}_\Omega\right)_b^2}.
\end{equation}

Using this condition, one can eliminate $\tilde\omega$ from the
expression for $\kappa_b$ in Eq.~(\ref{DoubleRoot}) and present it
in the alternative form:

\begin{equation}%
\label{DoubleRoot2} %
\kappa_b =
\frac{1}{16\left(\mbox{F}_\Omega\right)_b^2}\frac{\left[4\Omega\left(\mbox{F}_\Omega\right)_b^2\right]^2
- \left\{4(1 - \Omega) + \Omega^2\left[1 -
\left(\mbox{F}_\Omega\right)_b^2\right]\right\}^2}{4 - \Omega(4 -
\Omega)\left[1 - \left(\mbox{F}_\Omega\right)_b^2\right] -
4\Omega\left(\mbox{F}_\Omega\right)_b^2}.
\end{equation}

In two particular cases, $\Omega = 0$ and $\Omega = 1$, we have
\begin{equation}%
\label{DoubRt-01} %
\kappa_b|_{\Omega = 0} =
-\frac{1}{4\left(\mbox{F}_\Omega\right)_b^2}, \quad
\kappa_b|_{\Omega = 1} = -\frac{1 -
2\left(\mbox{F}_\Omega\right)_b^2
-15\left(\mbox{F}_\Omega\right)_b^4}{16\left(\mbox{F}_\Omega\right)_b^2\left[1
- \left(\mbox{F}_\Omega\right)_b^2\right]}.
\end{equation}

The normalized frequency, $\tilde\omega_b$, as per
Eq.~(\ref{DoubRtPart}), monotonically decreases with
$\mbox{F}_\Omega$ in the absence of the vorticity, i.e. when
$\Omega = 0$ [see dashed line in Fig.~\ref{f06}a) which
asymptotically approaches solid line 1 when
$\left(\mbox{F}_\Omega\right)_b \to 0$]. The dependence
$\left(\mbox{F}_\Omega\right)_b$ becomes formally non-monotonic
when $\Omega = 1$ [see another dashed line in Fig.~\ref{f06}a)
which approaches line 2 when $\left(\mbox{F}_\Omega\right)_b \to
0$]. However, the non-monotonic behavior of this function is just
an artefact of the deep-water approximation $\vert\kappa\vert \gg
1$ which is actually valid only when
$\left(\mbox{F}_\Omega\right)_b \ll 1$. The difference between the
corresponding dependences for the finite-depth water and
deep-water approximation is clearly seen in Fig.~\ref{f06}a) when
the Froude number increases.

The dependence (\ref{CondDoubleRoot}) can be also presented in the
form of the relationship between the critical Froude number
$\left(\mbox{F}_\Omega\right)_b$ and normalized wave period
$\tilde T_b$ similar to Eq.~(\ref{UvsT-ParamPos}); it is shown in
Fig.~\ref{f06}b) by dashed lines for $\Omega = 0$ and $\Omega = 1$
when it reduces to

\begin{equation}%
\label{CDRDV-lim}%
\tilde T_b|_{\Omega = 0} = 8\pi\left(\mbox{F}_\Omega\right)_b,
\quad \tilde T_b|_{\Omega = 1} =
16\pi\left(\mbox{F}_\Omega\right)_b \frac{\left[1 -
\left(\mbox{F}_\Omega\right)_b^2\right]^{3/2}}{\left[1 +
3\left(\mbox{F}_\Omega\right)_b^2\right]^2}.
\end{equation}%

As one can see from this figure, the deep-water approximation is
valid in the range of relatively small wave periods and small
Froude numbers. The deviation of dashed lines from solid lines
becomes noticeable when $\tilde T$ increases and approaches
$3\sqrt{3}\pi/2$ for $\Omega = 1$ or $\tilde T$ becomes greater
than 10 for $\Omega = 0$. In both cases of $\Omega = 0$ or $\Omega
= 1$ at small wave periods we have linear dependences
$\left(\mbox{F}_\Omega\right)_b = \tilde T/8\pi$ when $\Omega = 0$
and $\left(\mbox{F}_\Omega\right)_b = \tilde T/16\pi$ when $\Omega
= 1$. The linear dependence between the conventional Froude number
$\mbox{Fr} \equiv U_0/\sqrt{gh}$ and wave period for very short
gravity waves blocked by a uniform current with $\Omega = 0$ has
been derived in \cite{PRL09,NJP10}. Here we have found that for
the current with a constant vorticity similar linear dependence
between the `effective' Froude number $\mbox{F}_\Omega$ and wave
period takes place too, but with two times less gradient. However,
taking into account the relationship between the conventional
Froude number and `effective' Froude number [see the text after
Eq.~(\ref{DLDisp})], we obtain that when $\mbox{Fr} \to 0$ in both
case we have the same dependence $\mbox{Fr}_b = \tilde T/8\pi$.

\subsection{The shallow-water case}%
\label{Subsect3}%

For the sake of completeness, consider also another limiting case
$|\kappa| \ll 1$, $\kappa < 0$, which corresponds to the
shallow-water approximation. In this case the dispersion relation
(\ref{DLDisp}) reads:
\begin{equation}%
\label{SWDLDisp} %
\tilde\omega \approx -\frac{(2 - \Omega)\left(1 -
\mbox{F}_\Omega\right)}{2\mbox{F}_\Omega}\kappa\left[1 -
\frac{\left(2 - \Omega + \Omega\mbox{F}_\Omega\right)^2}{6(2 -
\Omega)^2\left(1 -
\mbox{F}_\Omega\right)}\kappa^2\right] + o(\kappa^3).%
\end{equation}%

From the condition of the wave blocking, $d\tilde\omega/d\kappa =
0$ we obtain:
\begin{equation}%
\label{SWNegRoot}%
\kappa_b = -\frac{\sqrt{2}(2 - \Omega)\sqrt{1 -
\left(\mbox{F}_\Omega\right)_b}}{2 - \Omega +
\Omega\left(\mbox{F}_\Omega\right)_b}, \quad \tilde\omega_b =
\frac{\sqrt{2}(2 - \Omega)^2\left[1 -
\left(\mbox{F}_\Omega\right)_b\right]^{3/2}}{3\left(\mbox{F}_\Omega\right)_b\left[2
- \Omega + \Omega\left(\mbox{F}_\Omega\right)_b\right]}.%
\end{equation}%

In the particular cases of uniform flow, $\Omega = 0$, and
constant vorticity flow, $\Omega = 1$, these expressions reduce to

\begin{equation}%
\label{SWNegRoot0}%
\Omega = 0: \quad \kappa_b = -\sqrt{2}\sqrt{1 -
\left(\mbox{F}_\Omega\right)_b}, \quad \tilde\omega_b =
\frac{2\sqrt{2}}{3\left(\mbox{F}_\Omega\right)_b}\left[1 -
\left(\mbox{F}_\Omega\right)_b\right]^{3/2};
\end{equation}
and
\begin{equation}%
\label{SWNegRoot1}%
\Omega = 1: \quad \kappa_b = -\frac{\sqrt{2}\sqrt{1 -
\left(\mbox{F}_\Omega\right)_b}}{1 +
\left(\mbox{F}_\Omega\right)_b}, \quad \tilde\omega_b =
\frac{\sqrt{2}}{3\left(\mbox{F}_\Omega\right)_b}\frac{\left[1 -
\left(\mbox{F}_\Omega\right)_b\right]^{3/2}}{1 +
\left(\mbox{F}_\Omega\right)_b}.%
\end{equation}

These dependences of $\tilde\omega_b$ on
$\left(\mbox{F}_\Omega\right)_b$ are shown in Fig.~\ref{f06}a) by
dotted lines. They can be also presented in the form of the
relationship between the critical Froude number and the
dimensionless period $\tilde T$ of a carrier wave:

\begin{equation}%
\label{SWTvsU1-lim}%
\tilde T_b|_{\Omega = 0} = \frac{3\pi\sqrt{2}}{2\left[1 -
\left(\mbox{F}_\Omega\right)_b\right]^{3/2}}, \quad \tilde
T_b|_{\Omega = 1} = \frac{3\pi\sqrt{2}}{2}\frac{\left[1 +
\left(\mbox{F}_\Omega\right)_b\right]^{3/2}}{1 -
\left(\mbox{F}_\Omega\right)_b}.
\end{equation}%
These dependences are shown in Fig.~\ref{f06}b) by dotted lines
for large values of $\tilde T$, as they are only asymptotically
valid at very large wave periods (i.e. at small wave numbers
$|\kappa| \ll 1$).

It is clearly seen from these formulae that when
$\left(\mbox{F}_\Omega\right)_b \to 1$, the wave period goes to
infinity, and when $\left(\mbox{F}_\Omega\right)_b$ becomes equal
to one, then waves of any period will be blocked out. Note,
however, when the `effective' Froude number
$\left(\mbox{F}_\Omega\right)_b \to 1$, then the conventional
Froude number $\left(\mbox{F}_\Omega\right)_b \to \infty$ [see the
text after Eq.~(\ref{DLDisp})]. This means that infinitely intense
surface current with linear vertical profile is required to block
out the wave of infinitely long period, whereas such waves can be
blocked out by a finite value of flow speed with a uniform
vertical profile. Physically this is quite understandable. Indeed,
as it is well known (see, e.g., \cite{Landau-Lifshitz-88}), the
wave motion induced by a surface wave decays with the depth. The
characteristic scale of decay depends on the wavelength, which, in
turn, depends through the dispersion relation on the wave period.
The larger the period, the larger is the characteristic vertical
scale of the wave motion. When $\Omega = 1$, the mean current
velocity linearly decreases with the water depth, therefore its
influence on the wave train gradually decreases and vanishes at
the bottom. In the meantime, if the flow vorticity is zero,
$\Omega = 0$, then the current uniformly affects a wave train at
any depth.

\section{Conclusion}
\label{Sect4}%

Thus, we have demonstrated that the blocking phenomenon may occur
for surface waves counter propagating with respect to gradually
varying flow in horizontal direction. The flow may be uniform in
depth or linearly varying. In the former case, waves of any
periods can be blocked at a certain critical speed of the current,
whereas in the latter case, the higher the wave period the higher
current surface speed is required to block out the wave.

We focused here on study of surface gravity waves only leaving
aside the effect of surface tension. With the surface tension the
problem becomes more complicated as gravity-capillary waves may
experience double reflection in two separate blocking points
\cite{Badulin-1983, Pokaz-Rozenb-1983, Trulsen-Mei-93}. We plan to
study this interesting phenomenon with and without main flow
vorticity in our next publication.

\acknowledgments This work was initiated when one of the authors
(Y.S.) visited the Laboratory J.-A. Dieudonn\'e of the University
of Nice in January--February 2011. Y.S. is highly appreciated the
Laboratory staff for the invitation and hospitality. The study was
supported by the Ministry of Education and Science of Russian
Federation, Project No 14.B37.21.0881.

\section{Appendix. Energy of surface waves of infinitesimal amplitude in the uniform flow}%
\label{Sect1}%

Following \cite{FabrStep-98}, consider the basic set of
hydrodynamic equations in the linear approximation for surface
gravity-capillary waves on the uniform flow whose velocity does
not depend on depth $U(z) \equiv U_0$ (see line 1 in
Fig.~\ref{f01}):

\begin{eqnarray}%
\frac{\prt u}{\prt t} + U_0\frac{\prt u}{\prt x} &=&
-\frac{1}{\rho}\frac{\prt p}{\prt x}, \label{A01} \\
\frac{\prt v}{\prt t} + U_0\frac{\prt v}{\prt x} &=& -\frac{1}{\rho}\frac{\prt p}{\prt z} - g, \label{A02} \\
\frac{\prt u}{\prt x} + \frac{\prt v}{\prt z} &=& 0, \label{A03}%
\end{eqnarray}%
where $\rho$ is the water density.

For the sake of generality we take into account the surface
tension effect which is cha\-racterized by the parameter $\sigma$
(if $\sigma = 0$ we obtain pure gravity waves). In the second part
of this work which will be published shortly we plan to consider
the blocking phenomenon for gravity-capillary waves and pure
capillary waves. Therefore, the results derived in this Appendix
will be used in our next pare too.

Augment the basic set of equations by the boundary conditions:
\begin{eqnarray}%
v(x, t) &=& 0, \quad \mbox{at } z = -h; \label{A04} \\
\frac{\prt \eta}{\prt t} + U_0\frac{\prt \eta}{\prt x} &=& v, \quad \mbox{at } z = 0; \label{A05} \\
p &=& P_a - \sigma \frac{\prt^2 \eta}{\prt x^2}, \quad \mbox{at } z = \eta, \label{A06}%
\end{eqnarray}%
where $P_a$ is the atmospheric pressure. The last boundary
condition (\ref{A06}) should be written at the water surface $z =
\eta$ and then only linear terms on wave amplitude $\eta$ should
be kept in consistency with the linear approximation.

The solution of this set can be presented in the form:
\begin{eqnarray}%
\eta(x,t) = \eta_0e^{i(\omega t - kx)}, \label{A07} \\
u(x,z,t) = \frac{\omega - U_0k}{\sinh{kh}}\eta_0\cosh{k(z + h)}e^{i(\omega t - kx)}, \label{A08} \\
v(x,z,t) = i\frac{\omega - U_0k}{\sinh{kh}}\eta_0\sinh{k(z + h)}e^{i(\omega t - kx)}, \label{A09} \\
p(x,z,t) = P_a - \rho g z + \rho \frac{\left(\omega -
U_0k\right)^2}{k\sinh{kh}} \eta_0\cosh{k(z + h)}e^{i(\omega t -
kx)}. \label{A10}%
\end{eqnarray}%

One can readily proof that solution (\ref{A07})--(\ref{A10})
automatically satisfy Eqs.~(\ref{A01})--(\ref{A03}) and boundary
conditions (\ref{A04})--(\ref{A05}), whereas substitution of
Eqs.~(\ref{A07}) and (\ref{A10}) into the boundary condition
(\ref{A06}) gives the dispersion relation (\ref{DispRel}) provided
that the flow is uniform in depth.

Using this solution, let us calculate the total energy of a
sinusoidal wave, the sum of the potential energy $P$ and kinetic
energy $K$: $E = P + K$.

The potential energy does not depend on a flow; it is determined
entirely by deflection of the free surface from the equilibrium
position $z = 0$:
\be%
\label{A11}%
P = \frac{\rho g}{2}\langle \eta^2\rangle +
\frac{\sigma}{2}\left\langle \left(\frac{\prt \eta}{\prt
x}\right)^2\right\rangle = \frac{\rho g}{4}\left(1 + \frac{\sigma
k^2}{\rho g}\right)\eta_0^2,
\ee%
where angular brackets stand for averaging over a wave period.

The kinetic energy of wave motion can be determined as the
difference between the kinetic energy of moving fluid with a wave
perturbation and without the perturbation:
\begin{equation}%
\label{A12} %
K = \frac{\rho}{2}\left\langle
\int\limits_{-h}^{\eta}\left[(u + U_0)^2 + v^2 -
U_0^2\right]\,dz\right\rangle = \frac{\rho}{4k}\frac{\omega^2 -
U_0^2k^2}{\tanh{kh}}\eta_0^2.%
\end{equation}%

Summing up $P$ and $K$, we obtain for the total wave energy (cf.
\cite{Dysthe2004})
\be%
\label{A13}%
E = \frac{\rho \eta_0^2}{4k\tanh{kh}}\left[\omega^2 - U_0^2k^2 +
\left(1 + \frac{\sigma k^2}{\rho g}\right)gk\tanh{kh}\right] =
\frac{\rho \eta_0^2}{2k\tanh{kh}}\omega\left(\omega - U_0k\right).
\ee%

Equation (\ref{A13}) shows that the wave energy is proportional
both to the frequency $\omega$ and relative frequency $(\omega -
U_0 k)$, which is the frequency in the frame co-moving with the
fluid. Thus, as the frequency $\omega$ is always positive in the
immovable coordinate system, then we see that the energy may be
negative if the relative frequency is negative; in this case we
deal with the so called negative energy waves (NEWs) (see
\cite{Nezlin-76, OstrRybTsim-86, StepFabr-89, FabrStep-98}).

Substituting here the expression for the frequency
\begin{equation}%
\label{DispGravCap}%
\omega = U_0k - \sqrt{\left(gk + \sigma k^3/\rho\right)\tanh{kh}}%
\end{equation}
of gravity-capillary waves counter-current propagating ($k < 0$)
on the uniform flow \cite{Choi, Wahlen}, we obtain:
\be%
\label{A14}%
E = \frac{\rho g}{2}\eta_0^2\left(1 + \frac{\sigma k^2}{\rho
g}\right)\left[1 - \frac{U_0}{\sqrt{gh}}\sqrt{\frac{kh}{\left(1 +
\sigma k^2/\rho g\right)\tanh{kh}}}\;\right].
\ee%

In the dimensionless variables as defined in the text, the wave
energy reads:
\be%
\label{A14}%
E = \frac{\rho g}{2}\eta_0^2\left(1 +
\mbox{S}\kappa^2\right)\left[1 -
\mbox{Fr}\sqrt{\frac{\kappa}{\left(1 +
\mbox{S}\kappa^2\right)\tanh{\kappa}}}\;\right],
\ee%
where $\mbox{S} = \sigma/\left(\rho gh^2\right)$ is the parameters
which measures the strength of the capillary effect relative to
the gravity effect.

As follows from Eq.~(\ref{A14}), the energy becomes negative when
\be%
\label{A15}%
\mbox{Fr} > \sqrt{\frac{\left(1 +
\mbox{S}\kappa^2\right)\tanh{\kappa}}{\kappa}}.
\ee%

For pure gravity waves without surface tension ($\mbox{S} = 0$)
this expression gives the same threshold for appearing of NEWs as
Eq.~(\ref{kappa0}) derived from the dispersion relation
(\ref{DLDisp}) with $\Omega = 0$.

As has been explained in \cite{StepFabr-89, FabrStep-98}, the
concept of negative energy waves helps to find ``potentially
unstable waves'' which may grow if there is a mechanism taking
away their energy. In the absence of such mechanism, they are
neutrally stable. From the physical point of view, NEWs are waves
moving slower than the fluid layer. In the ``laboratory''
coordinate frame where fluid flows to the right (see
Fig.~\ref{f01}) with the velocity $U_0$, those waves move slower
than the fluid, whose phase velocity being directed originally to
the left ($k < 0$, $c_{ph} \equiv \omega/k < 0$) are pulled down
by the flow and propagate eventually to the right. As follows from
the dispersion relation for the uniformly moving fluid
(\ref{DispGravCap}), this occurs when
\begin{equation}%
\label{A16} %
U_0 > \sqrt{\frac{g\tanh{kh}}{k}\left(1 +
\frac{\sigma}{\rho g} k^2\right)},%
\end{equation}%
where the right-hand side of this inequality represents the phase
speed of gravity-capillary waves on a calm water. This is nothing
but the dimensional form of Eq.~(\ref{A15}).

\end{document}